\documentclass{elsart}

\usepackage{amssymb}

\newcommand{\pom}{\ensuremath{{I\!\!P}\/\/ }}

\newcommand{\Journal}[4]{ {#1} {\bf #2}, {#3} (#4)}

\newcommand{\EPA}{Eur.\ Phys.\ J.\ A}
\newcommand{\EPC}{Eur.\ Phys.\ J.\ C}

\newcommand{\JPG}{J.\ Phys.\ G }

\newcommand{\NPA}{Nucl.\ Phys.\ A}
\newcommand{\NPB}{Nucl.\ Phys.\ B}

\newcommand{\PiNL}{$\Pi$N Newslett.\ }

\newcommand{\PLB}{Phys.\ Lett.\ B}

\newcommand{\SNP}{Sov.\ J.\ Nucl.\ Phys.\ }

\newcommand{\ZPC}{Z.\ Phys.\ C}

\begin{document}

\begin{frontmatter}



\title{
The Structure of the Pion and Nucleon, 
and Leading Neutron Production at HERA
}


\author{Garry Levman}

\address{Department of Physics, University of Toronto,
Toronto ON M5S 1A7, Canada}
\ead{levman@physics.utoronto.ca}
\ead[url]{http://www.physics.utoronto.ca/\~\,levman}

\begin{abstract}
Attention is paid to recent results 
from the ZEUS Collaboration on the photo- and electro-
production of leading neutrons in $e^+ p$ collisions
at HERA.
Some implications for the
structure of the pion and the nucleon are discussed.
\end{abstract}

\end{frontmatter}

\section{Introduction}

The ZEUS Collaboration has published data on
leading neutron production in neutral current $e^+ p$ collisions 
at HERA\cite{zfnc3}.
The data have been anticipated because of the light
they cast on the structure of the pion and nucleon.

In \cite{zfnc3} the ZEUS Collaboration uses
an {\it effective flux method}\cite{ringberg}
to determine
the photon-pion total cross section at high W,
and 
the deep inelastic structure function of the pion,
$F_2^{\pi}(x,Q^2)$\footnote[1]{
The notation used follows \cite{zfnc3}. 
W is a center of momentum frame energy.
$x$ is a Bjorken scaling variable.
$Q^2$ is the virtuality of the exchanged photon.}.
In this note, the method is described; its assumptions and
inherent difficulties are discussed; and the problem of error
is confronted.
Implications of the ZEUS data
are examined critically
for the role of meson exchange in the production of
leading neutrons and for the structure of the pion.

\section{Meson Exchange Picture of Leading Baryon Production}

Leading baryons are those produced with small
transverse momentum, $p_T$, and carrying a large fraction
$x_L$, of the incoming proton's energy. 
They are thought to represent the conserved baryon current
because the
squared 4-momentum transfer, $t$, between the incoming proton 
and outgoing baryon is small(\cite{zfnc3}, Fig.~1).

The production data for leading baryons can be
understood in terms of meson exchanges.  
Consider the specific case $\gamma p\rightarrow Xn$.
The cross section for the production
of a leading neutron is given by a sum over all possible
meson exchanges between the photon and proton which lead to
an outgoing neutron at the proton vertex(\cite{zfnc3}, Fig.~1):  
\begin{equation}
\sigma_{\gamma p\rightarrow X n}(x_L,t,W^2) \sim 
        a_{\gamma,p}(t)
        \sum_i
        f_{i/p}(x_L,t)\sigma_{\gamma i}\left( (1-x_L)W^2,t\right)
\label{exchange}
\end{equation}
where
$i = \pi, \rho$, or $a_2$ meson;
$\sigma_{\gamma i}$ is the total
$\gamma i$ cross section;
$f_{i/p}$ is the splitting function for
$p\rightarrow n$ via $i$ exchange; and
$a_{\gamma,p}$ is a form factor which accounts
for rescattering of the neutron (absorption) due
to the finite size of the projectile. 
Projectiles other than the photon, and produced
baryons other than the neutron are 
treated similarly. For example, in the reaction
$p p\rightarrow X p$ the $\omega$ and $f_2$ mesons, and
the pomeron $\pom$ also contribute.
There are complications.
The direct production of baryons which decay, such as 
$\Delta\rightarrow n\pi$,
leads to the indirect production of nucleons;
there can be two-meson exchange, and
different exchanges can interfere;
the form factor, $a_{\gamma ,p}$, can depend on $x_L$ 
and perhaps on the exchanged meson $i$.
The simple factorization assumed
in Eqn.~\ref{exchange} is only an approximation.

\section{Leading Neutron Production and Pion Exchange}

In hadro-production experiments
(projectile=$p,\pi$) leading neutron production
is believed to be dominated by pion exchange.
The spectrum of leading neutrons in $ pp $ 
collisions at Fermilab and the ISR is fairly well 
represented 
by a single term (one-meson-exchange) of the form
\begin{equation}
\sigma_{pp\rightarrow Xn}=
   \frac{1}{4\pi} \frac{g^2}{4\pi}
                         \frac{-2t}{(t-m^2_{\pi})^2}
                          (1-x_L)^{1-2t}
   \sigma_{\pi p}\left( (1-x_L)W^2,t=0\right)
\label{ope}
\end{equation}
where  $m_{\pi}$ is the mass of the pion, 
$g^2/4\pi = 14.5$,
and $\sigma_{\pi p}$ is the 
measured, on-mass-shell, $\pi p$ total cross section.
The splitting function
\begin{equation}
f_{eff}(x_L,t) =  \frac{1}{4\pi} \frac{g^2}{4\pi}
                         \frac{-2t}{(t-m^2_{\pi})^2}
                          (1-x_L)^{1-2t}
\label{feff}
\end{equation}
is very close to that expected from pure one-pion exchange.
The result is simple, but surprising. Note in Eqn.~\ref{feff}
the absence of
\begin{itemize}
\item an absorptive factor $a(t)$;
\item an off-mass-shell correction,
      that is, a $t$ dependence to $\sigma(\pi p)$;
\item contributions from $\rho$
      and $a_2$ exchange, and $\Delta$ production.
\end{itemize}
The self-consistency of the meson exchange picture
requires the presence of such `backgrounds'.
In addition, the value, 14.5,  of the $p\pi^0 p$ coupling
constant is larger than  recent estimates\cite{deswart,ericson} 
which suggest that $g^2/4\pi$ lies in the range (13.5--14.1).

The ZEUS Collaboration has reviewed (see \cite{zfnc3}, \S10.1)
the expected backgrounds to one-pion exchange
which are known both from experimental
measurements (eg.\ $\Delta$ production) and from 
theoretical calculations (eg.\ absorption).
The effects are large, roughly 20-30\%,
apparently confounding the experimental observation
embodied in Eqns.~\ref{ope} and~\ref{feff};
however,
\begin{itemize}
\item $\rho$, $a_2$ exchange and $\Delta$ production
{\it increase} neutron production, while
absorptive effects { \it decrease} neutron production
\item absorption preferentially removes
      $\rho$ and $a_2$ compared to $\pi$ exchange because
      $a(t)$ decreases with increasing $t$ and the 
      contribution of the higher mass mesons increases
      relative to the pion at high~$t$.
\end{itemize}

The conclusion to draw is that {simple one-pion exchange
agrees well with the hadro-production data because of a
fortuitous near cancellation of absorptive effects and
background contributions}. The large value of the
effective coupling constant (14.5) implies that the
cancellation is not perfect, and that there is a
residual $\lesssim 5$\% contribution to the normalization.
$f_{eff}$ is not the true pion flux, but
rather an effective pion flux in the presence of
competing and `compensating' processes.

\section{The Effective Flux}

The effective flux of pions in the proton 
(effective splitting function) can be defined 
by experiment as the deconvolution
of the {measured} semi-inclusive differential cross section 
for $p p\rightarrow Xn$
and the {measured} $\pi p$ total cross section,
\begin{equation}
f_{eff}(x_L, t) \equiv \frac{d\sigma_{pp\rightarrow Xn}}{dx_L dt} / \sigma_{\pi p}
\label{definition}
\end{equation}
where deconvolution is written as division.
By construction and definition $f_{eff}$ corrects 
one-pion exchange in $pp$ collisions for 
$\rho$ and $a_2$ exchange, $\Delta$ production, 
rescattering and
off-mass-shell effects, etc.
From Eqn.~\ref{exchange} it is seen that a `model'
for $f_{eff}$ is 
\[ 
f_{eff} \sim  a_{p,p}(t)\sum_i
        f_{i/p}(x_L,t)\sigma_{i p}\left( (1-x_L)W^2,t\right)/
        \sigma_{\pi p}\left( (1-x_L)W^2,t=0\right)
\]
It is important to emphasize:
experiments find that a convenient parameterization of $f_{eff}$
is given by Eqn.~\ref{feff}.
Because of the universal
behavior of hadronic cross sections
$f_{eff}$ is also expected
to work for ${\pi p}$ and $\gamma p$ collisions.

\section{The Structure of the Pion and the Effective Flux}

It has been proposed\cite{holtmann,kopeliovich}
to measure the structure 
function of the pion $F_2^{\pi}$ at HERA using
virtual $\gamma^* \pi$ collisions 
assuming pion exchange
dominates the interaction 
$\gamma^* p\rightarrow X n$, just as it does in hadro-production. 

A vital consistency check for any extraction of
$F_2^{\pi}$ using the electro-production of leading 
neutrons is the demonstration that the
photo-production of leading neutrons is in accord with
the hadro-production measurements. The importance of the
photo-production data arises because:
\begin{itemize}
\item $W$ is the single relevant leptonic variable ($Q^2=0$).
\item the $W$ dependence of measured hadronic 
      cross sections closely follows a simple 
      universal behavior;
      at high $W$ the hadronic total cross sections behave as
      a power law $W^{2(\alpha_{\pom}-1)}$, where $\alpha_{\pom}$
      is a constant independent of the projectile and target.
\item the photo-production ($Q^2<0.02$ GeV$^2$) and 
      transition region ($0.1 < Q^2 < 0.74$ GeV$^2$)
      data from HERA are in good 
      agreement with vector dominance in which the photon behaves like an
      exchanged vector meson
      and the collision is hadronic\cite{zphenlowq,zsigtot}.
\end{itemize}

In contrast to photo-production, 
the situation in electro-production
is less well understood.
The additional degree of freedom quantified by $Q^2$
adds complication. Rescattering, background composition and
levels, and factorization properties can vary with $Q^2$.
Moreover, in deep inelastic scattering
($Q^2\gtrsim 4$ GeV$^2$) the chief interest lies
in {measuring} the $W$ (that is $x$, $x\approx Q^2/W^2$ at low $x$)
dependence of the cross section.
In summary, a prerequisite for a {believable}
determination of $F_2^{\pi}$ using high $Q^2$ electro-production
data is a consistent determination of $\sigma_{\gamma\pi}$
using photo-production data. 

A standard technique for determining $\sigma(\gamma\pi)$
involves fitting the observed differential cross section
for $\gamma p\rightarrow X n$ while simultaneously correcting
for background exchanges, Delta production, absorption,
and off-mass-shell effects\cite{zakharov}.
This procedure has the 
important advantage of providing an estimate of the
size of the contributing sub-processes and giving 
a transparent estimate of the error. It has the 
disadvantage that there is a large number of effects
to be accounted for, with a corresponding large 
degree of freedom. 
The parameters must be guessed, fit, or
fixed by experiment.

The effective flux is used instead to directly
determine $\sigma(\gamma\pi)$ without determining
the individual sub-processes. This has the advantages
of simplicity and directness, but the disadvantage
that no information is obtained about the relative
importance of the various processes which contribute.
The error is difficult to estimate (see \S\ref{discussion})
because the amount of pion exchange present is not
unambiguously determined. If pion exchange does
not contribute, the result, although well
defined, is meaningless.

The arguments which follow (in 
\S\ref{sigpi} and \S\ref{f2pi})
give the essence of what obtains from the
ZEUS measurement.

\subsection{$\sigma(\gamma\pi)$ at high $W$}
\label{sigpi}

The $\gamma\pi$ total cross section
is obtained from the photo-production data using
\[
\sigma (\gamma\pi) =\frac{d\sigma (\gamma p\rightarrow Xn)/d x_L}{f_{eff}}
\]
where the variable $t$ has been integrated out in the numerator.
Substitution for $f_{eff}$ from its definition in
Eqn.~\ref{definition} shows that $\sigma(\gamma\pi)$ can 
be determined from (double) ratios
\begin{equation}
\frac{\sigma (\gamma\pi)}{\sigma (\gamma p)} 
    =
 \frac{\displaystyle\frac{d\sigma (\gamma p\rightarrow Xn)/dx_L}{\sigma(\gamma p)}}
      {\displaystyle\frac{d\sigma (p p\rightarrow Xn)/dx_L}{\sigma(p p)}}
  \cdot\frac{\sigma (\pi p)}{\sigma ( p p)}                  
\label{double}
\end{equation}
of measured quantities.

The ZEUS Collaboration has observed that the relative rate
of neutron production in photo-production at HERA is
{\it half} that of $pp$ collisions.
It follows from Eqn.~\ref{double} that
$\sigma (\gamma\pi)/\sigma (\gamma p)$ is half
$\sigma (\pi p)/\sigma ( p p)$.
Therefore, as ZEUS deduces directly,
\[
\sigma (\gamma\pi) \simeq \sigma (\gamma p)/3
\]
rather than two-thirds as expected
from Regge factorization or the 
counting of valence quarks
(the Additive Quark Model).

An important consistency check,
which the ZEUS Collaboration
has performed (\cite{zfnc3}, Fig.~6 \& 8),
is that the neutron energy and angular distributions 
are described by $f_{eff}$. 

\subsection{$F^{\pi}_2$ at low $x$}
\label{f2pi}

The double ratio (Eqn.~\ref{double})
is robust for hadro- and photo-production; 
however, for electro-production,
absorptive rescattering 
decreases as $Q^2$ increases, 
as discussed in \cite{nszak} and \cite{alesio}
(see also \cite{zfnc3}, Fig.~9).
At low $Q^2$ ($Q^2\lesssim 1$ GeV$^2$), $F_2^{\pi}$ and 
$\sigma (\gamma^*\pi)$ are related by
\begin{equation}
F_2^{\pi}\simeq\frac{Q^2}{4\pi^2\alpha}\sigma (\gamma^*\pi)
\label{php}
\end{equation}
which, in analogy with $\rho$ dominance in
$\gamma p$ interactions\cite{zphenlowq},
can be written as
\begin{equation}
F_2^{\pi}\simeq\frac{Q^2}{4\pi^2\alpha}\frac{m_{\rho}^2}{m_{\rho}^2+Q^2}
         \sigma (\gamma \pi)
\label{bpc}
\end{equation}
where $m_{\rho}$ is the mass of the $\rho$ meson. 

Equations~\ref{php} and~\ref{bpc} connect the
photo-production $\gamma\pi$ cross section
to the transition region $\gamma^*\pi$ cross section;
together with the corresponding equations for $\gamma^{(*)} p$
interactions,
\begin{equation}
\frac{ \sigma(\gamma^* \pi)}{\sigma(\gamma \pi)}\simeq 
\frac{ \sigma(\gamma^* p)}{\sigma(\gamma p)}
\label{ratio}
\end{equation}
obtains. 
In the kinematic region $0<Q^2<4$ GeV$^2$ the ratio of
neutron production (i.e. tagged divided by all) increases. 
To maintain Eqn.~\ref{ratio} an
absorptive correction needs to be applied.
Rescattering {decreases}, and 
the effective flux must {increase}
correspondingly in order to maintain
consistency between
photo-production, the transition region, and 
deep inelastic scattering.
The ZEUS data require for deep inelastic
scattering (DIS)
$f_{eff}\rightarrow f^{\rm DIS}_{eff}=1.3f_{eff}$.

The measurements 
show that 2.5\% of deep inelastic scattering
events, independent of $x$ and $Q^2$, have
a leading neutron with $0.64<x_L<0.82$ and
$p_T<0.66x_L$ GeV (\cite{zfnc3}, Fig.~10). 
The constancy of the ratio of neutron production in DIS
as a function of $x$ and $Q^2$ implies that
$F_2^{\pi}(x,Q^2)=kF_2^p\left((1-x_L)x,Q^2\right)=kF_2^p(0.27x,Q^2)$
with a constant of proportionality $k$
given by 
\[
k=\frac{0.025}{\int\int f^{\rm DIS}_{eff}(x_L,t) dx_L dt} = 0.25
\]
where the integration is over the $x_L$ and $t(p_T)$
range covered by the ZEUS measurement.
Because
$F_2^p(x)\propto (1/x)^{\lambda}$
with $\lambda\approx 0.2$\cite{zphenlowq} 
\[
F_2^{\pi}(x,Q^2)\approx F_2^p(x,Q^2)/3
\]

\section{Discussion}
\label{discussion}

For the determination of $F_2^{\pi}$ the effective flux method requires 
little theoretical input. Only the general concepts
of meson exchange theory are needed. The data,
hadro-production, photo-production and electro-production,
determine everything. Note especially that 
\begin{itemize}
\item  one-pion-exchange is not assumed. It is only assumed
   that pion exchange dominates neutron production.
\item the contribution of backgrounds is not neglected.
    The method accounts for $\rho$ and $a_2$ exchange and 
    for indirect neutron production through $\Delta$ 
    production and decay.
\item the exact composition and relative contribution of the 
     backgrounds is unimportant.
\item  absorptive effects are measured.
\item the consistency of the photo-production ($Q^2< 0.02$ GeV$^2$),
   transition region ($0.1< Q^2 < 0.74$ GeV$^2$), and DIS ($Q^2>4$ GeV$^2$)
   analyses requires and defines an absorptive correction.
\end{itemize}
If absorption approximately compensates for
background exchanges in hadro- and photo-production, as the data suggest,
then the 30\% rise in neutron ratio
between photo-production and DIS is a measure of the relative 
contribution of the backgrounds.
  
The effective flux gives a normalization
of $\sigma(\gamma \pi)$ which differs by a factor of two
from that expected by Regge factorization or
valence quark counting.
The disagreement follows directly from
the experimental observation that the relative rate of neutrons
in photo-production is half that observed in 
hadro-production.
Can $f_{eff}$, which is determined by
$pp$ collisions, be wrong by a factor of two when used for
photo-production?

Background exchanges and absorptive effects are of the same
magnitude,
but contribute with opposite signs. 
The ZEUS leading neutron data
show that the relative size of absorptive effects 
is approximately 20-30\%. 
Background processes must contribute at approximately this level,
in agreement with estimates from phenomenological
studies (reviewed in\cite{zfnc3}, \S10.1).
With this in mind, a conservative estimate for the error on using
$f_{eff}$ for photo-production is obtained by assuming 
that pion exchange contributes $\gtrsim 50$\% and 
that the backgrounds change by $\lesssim 50$\%
on moving from $pp$ to $\gamma p$ collisions.
The change in $f_{eff}$ is then $\lesssim 25$\%.
Suppose that the low neutron rate is not due
to a small photon-pion cross section, but rather due
to changes in absorption or backgrounds. Then 
absorption must increase or backgrounds decrease
strongly in the photon-proton system compared to the
proton-proton system.

If the meson exchange picture underestimates the
normalization of $\sigma(\gamma\pi)$ by a factor of two,
the theory is wrong, misapplied, or
there is a significant missing ingredient. 
Then
{one cannot extract with any confidence the $x$ and
$Q^2$ behavior of $F_2^{\pi}$}. 

If the photo-production result is disbelieved,
the ZEUS data can be taken as evidence that
meson exchange plays {\it little} role in the production of 
leading nucleons at HERA.
One then argues that the ZEUS measurement
is merely a reflection 
of $F_2^p$, arising because of 
factorization (limiting fragmentation) at
the proton vertex and baryon conservation.

On the other hand, the experimental evidence for the meson exchange
picture is good (see references in \cite{zfnc3}).
One can accept the picture and the normalization
of $\sigma(\gamma\pi)$ and $F_2^{\pi}$ implied by the ZEUS
data. 
In this case some conjectures can be make:
\begin{itemize}
\item the $x$ dependence of $F_2$ for all hadrons is similar
      at low $x$ and is determined mainly by the QCD
      evolution equations, only weakly by the valence structure
      of the hadron.
\item the number of partons at low $x$ in the pion is 1/3 that 
      of the proton;
      since the charged radius of the pion is 2/3 the proton's,
      the {\it volume} density of partons in the pion is approximately 
      the same as in the proton.
\item there is a large probability for the proton to be found in
      a meson-nucleon or meson-Delta Fock state (the nucleon
      within the nucleon); the proton is 
      a loosely bound meson-nucleon system composed of infinitely
      bound partons.
\item the quark-antiquark sea of a hadron is generated mainly
      by valence-valence interactions 
      (three for the proton and one for the pion),
      and not by self interactions.
\end{itemize}
Moreover, there is a significant violation of the
quark counting rules and Regge factorization. 

\section {Conclusions}

In the context of the meson exchange picture of leading
baryon production, the leading neutron data from ZEUS, 
taken together with the hadro-production data from
Fermilab and the ISR, make
a statement about the normalization of $\sigma(\gamma\pi)$.
With the assumption that the $x$ and $Q^2$ dependence of the difference
between pion exchange and the backgrounds is weak, $F_2^{\pi}$
can be extracted from the data. 
The normalization is the firmer of the determinations.
If the normalization fixed by the 
photo-production data is incorrect, then an 
extraction of $F_2^{\pi}$ is doubtful
since the self-consistency of the picture
is questionable. In particular, the $x$ and $Q^2$ dependence extracted
from the data is questionable because
the high $Q^2$ region is {less} well understood than 
photo-production,
both experimentally and phenomenologically.

If the determination of $F_2^{\pi}$ using the effective flux method
is correct, then the ZEUS data have important implications for 
the structure of the nucleon and the pion.

\section*{Acknowledgments}

I thank the ZEUS Collaboration for many useful discussions
and for its efforts in publishing the inclusive leading neutron data.
I thank John Martin and Malcolm Derrick for helpful comments.
\newline

\end{document}